\documentstyle[prl,aps,floats,graphics,twocolumn]{revtex}

\author{J. Houdayer and O. C. Martin}
\title{Renormalization for Discrete Optimization}
\address{Laboratoire de Physique Th\'eorique et Mod\`eles Statistiques,\\
b\^at. 100, Universit\'e Paris-Sud, F--91405 Orsay, France.}
\date{\today}
\draft

\begin{document}
\maketitle

\begin{abstract}
The renormalization group has proven to be 
a very powerful tool in physics
for treating systems with many length scales. Here 
we show how it can be adapted to provide a new 
class of algorithms for discrete optimization.
The heart of our method uses renormalization and recursion, and
these processes are embedded in a genetic algorithm.
The system is self-consistently optimized on
all scales, leading to a high probability of finding
the ground state configuration. To demonstrate the generality of
such an approach, we perform
tests on traveling salesman and spin glass problems. The
results show that our ``genetic renormalization algorithm''
is extremely powerful.
\end{abstract}


The study of disordered systems is an active and challenging
subject~\cite{Young98}, and in many cases some of the most basic
consequences of randomness remain subject to controversy. Given that
numerical calculations of ground state properties can shed light on these
issues, it is not surprising that more and more such calculations are being
performed~\cite{Rieger98}. Our goal here is to introduce and test a new
general purpose approach for finding ground states in disordered and
frustrated systems. In this letter we illustrate its use on the traveling
salesman problem and on the spin glass problem, showing that the ground
states are found with a high probability. More generally, our novel approach
should be very useful for many classes of discrete optimization problems and
is thus of major interdisciplinary interest.

Although it is often claimed that physical insight into disordered systems
should lead to improved optimization algorithms, thus far, there has been
very little substance to uphold this view. Aside from simulated
annealing~\cite{KirkpatrickGelatt83} and generalizations
thereof~\cite{MobiusNeklioudov97}, physics inspired ideas, ranging from
replica symmetry breaking to energy landscapes, have had little impact on
practical algorithmic developments in optimization. Nevertheless, several
ideas from physics seemed promising, including
renormalization~\cite{KawashimaSuzuki92} and hierarchical
constructions~\cite{YoshiyukiYoshiki95}. Perhaps, the impact of these
attempts has been minor because the resulting algorithms were not
sufficiently powerful to be competitive with the state of the art. In our
work, we have found that by carefully combining some of these ideas, namely
renormalization and recursion, {\it and} by embedding them in a genetic
algorithm approach, highly effective algorithms could be achieved. We thus
believe that the essence of the renormalization group can be fruitfully
applied to discrete optimization, and we expect the use of this type of
algorithm to become widespread in the near future.

Let us begin by sketching some of the standard approaches for tackling hard
discrete optimization problems~\cite{PapadimitriouSteiglitz82}. For such
problems, it is believed that there are no fast algorithms for finding the
optimum, so much effort has concentrated on the goal of quickly obtaining
``good'' near-optimum solutions by heuristic means. One of the simplest
heuristic algorithms is local search~\cite{AartsLenstra97} in which a few
variables are changed at a time in the search for lower energy
configurations. This heuristic and numerous generalizations thereof such as
simulated annealing~\cite{KirkpatrickGelatt83} optimize very effectively on
small scales, that is on scales involving a small number of variables, but
breakdown for the larger scales that require the modification of many
variables simultaneously. To tackle these large scales {\it directly},
genetic algorithms~\cite{Goldberg89} use a ``crossing'' procedure which
takes two good configurations (parents) and generates a child which combines
large parts of its parents. A population of configurations is evolved from
one generation to the next using these crossings followed by a selection of
the best children. Unfortunately, this approach does not work well in
practice because it is very difficult to take two good parents and cross
them to make a child which is as good as them. This is the major bottleneck
of genetic algorithms and is responsible for their limited use. For an
optimization scheme to overcome these difficulties, it must {\it explicitly}
treat all the scales in the problem simultaneously, the different scales
being tightly coupled. To implement such a treatment, we rely on ideas from
the renormalization group, the physicist's favorite tool for treating
problems with many scales~\cite{Ma76}. Our approach is based on embedding
renormalization and recursion within a genetic algorithm, leading to what we
call a ``genetic renormalization algorithm'' (GRA). To best understand the
working of this approach, we now show how we have implemented it in two
specific cases, the traveling salesman and the spin glass problems.

{\it The traveling salesman problem (TSP)} --- This routing problem is
motivated by applications in the telecommunication and transportation
industries. Given $N$ cities and their mutual distances, one is to find the
shortest closed path (tour) visiting each of the cities exactly
once~\cite{PapadimitriouSteiglitz82}. In genetic algorithms, one takes two
parents (good tours) from a population and finds the sub-paths they have in
common. Then a child is built by reconnecting those sub-paths, either
randomly or by using parts belonging to the parents if possible; ultimately,
these connections are not very good and lead to a child which is less fit
than its parents.

In our approach, instead of creating children as described, we {\it
engineer} new configurations from sub-paths that are frequently shared in
the population. In practice we pick $k$ ``parents'' at random and determine
their common sub-paths: these form the {\it patterns} which we select before
engineering the child. Then we wish to find the very {\it best} child which
is compatible with these patterns. (This child should thus be at least as
good as its best parent.) For this new problem, each sub-path is replaced by
its two end cities and one bond which connects them; together with the
cities which do not belong to any of the patterns, this defines a new,
``renormalized'', TSP with fewer cities. Note that in this new TSP, we have
removed all the cities inside the selected sub-paths, and have ``frozen-in''
bonds to connect their end-points; since we force these bonds to be in the
tours, the renormalized problem is really a constrained TSP. The distance
between two cities is the same as in the non-renormalized problem if they
are not connected by a frozen bond, otherwise their distance is given by the
length of the sub-path associated with the frozen bond. If this reduced
problem is small enough, it can be solved by direct enumeration. Otherwise,
we ``open up the Russian doll'' and solve this renormalized problem
recursively! Since each parent is compatible with the selected patterns,
each of them corresponds to a legal tour for the renormalized problem. Thus
we can use these tours in the first generation of the recursive call of GRA:
this way none of the information contained in the tours is lost.

How does one choose the number of parents, $k$? Clearly, the tour parts that
are shared by all $k$ parents decrease as $k$ grows and the child becomes
less and less constrained. Increasing $k$ then has the effect of improving
the best possible child but also of making the corresponding search more
difficult, so the choice of $k$ results from a compromise. Genetic
algorithms being biologically motivated, the choice $k=2$ may seem natural,
but it need not be optimal and empirically we find it not to be. We do not
claim to be the first to propose the use of more than two parents
\cite{MuhlenbeinGeorgesSchleuter88},
but in previous proposals, the performance turned out to be lackluster.
The reason is that
they did not include the two essential ingredients: (i) a selection
of patterns; (ii) a search for the best child consistent with the given
patterns.

A bird's eye view of our algorithm is as follows. We start with a population
of $M$ randomly generated tours; a simplified version of the
Lin-Kernighan~\cite{LinKernighan73} local search algorithm is applied to
these tours which form the first generation. To obtain the next generation,
we first produce by recursion as many children as there are parents; then
the local search improvement is applied to these children; finally,
duplications among the children and children which present no improvement
over their worst parent are eliminated. The next generation consists of the
children remaining. The algorithm terminates when there is only one
individual left.

If the local search is taken as given (and we are not concerned here about
its detailed implementation), our algorithm has two parameters, the number
$M$ of tours used in the population and $k$ the number of parents of a
child. In our numerical experiments for the TSP, we have chosen $M=50$ for
the top-most level where we treat the initial TSP instance, and $M=8$ for
the inner levels where renormalized instances are treated. Of course, other
choices are possible, but we have not explored them much. Let us just note
that it is desirable to have $M$ large enough to have plenty of diversity in
the patterns which will be selected, thereby increasing one's chance of
finding the ground state. However, there is a high computational cost for
doing this, as each level of the recursion increases the CPU time
multiplicatively. Thus the best strategy would probably be to have $M$
decrease with the level of the recursion. Concerning the choice of the
parameter $k$, a similar compromise has to be reached. The best quality
solutions would be obtained with large $k$, but this would lead to many
levels of recursion and thus to very long computation times. In practice, we
increase $k$ dynamically until of the current number of bonds to be found,
at least a threshold fraction of 10~\% remains unfrozen at this step. This
ensures that the renormalization does not reduce the problem size too
dramatically, allowing good solutions to be found. For the instances we
considered, nearly all values of $k$ were between $2$ and $6$, with $5$
being the most probable value.

\begin{table}
\begin{center}
\begin{tabular}{lrcccr}
instance & $\Delta_{LK}$ & $\tau_{LK}$ & $\Delta_{GRA}$
& $\tau_{GRA}/\tau_{LK}$ & $P_{GRA}$\\
\hline
pcb442 & 1.9 \% & 0.09 & 0 \% & 2442 & 100 \%\\
rat783 & 2.0 \% & 0.19 & 0 \% & 2923  & 100 \%\\
fl1577 & 15.4 \% & 0.63 & 0.0022 \% & 3805 & 80 \%\\
pr2392 & 2.7 \% & 0.98 & 0.0056 \% & 5278 & 20 \%\\
rl5915 & 3.6 \% & 3.94 & 0.013 \% & 8202 & 0 \%\\
\end{tabular}
\end{center}
\caption{Tests on 5 instances from
TSPLIB; the number in the name of an instance represents its number of
cities. $\Delta_{LK}$ and $\Delta_{GRA}$ are the relative differences
between the length found by the corresponding algorithm and the optimum,
$\tau_{LK}$ and $\tau_{GRA}$ are the CPU times in seconds to treat one
instance, and $P_{GRA}$ represents the probability for GRA to find the
optimum. Data for the GRA have been averaged over 10 runs.}
\label{TableTsp}
\end{table}

How well does the method work? For the TSP, it is standard practice to test
heuristics on problems from the TSPLIB library~\cite{Reinelt91}. We have
tested our algorithm on 5 problems of that library for which the exact
optima are known. As can be seen in Table~\ref{TableTsp}, the improvement
over the local search is impressive (we use a DEC-$\alpha$-500 work-station
to treat these instances). Still better results could be obtained by
improving the local search part. Several other groups
(see~\cite{MobiusNeklioudov97} and chapter 7 in~\cite{AartsLenstra97}) have
fine-tuned their Lin-Kernighan algorithm both for speed and for quality. In
spite of the fact that our version of LK is far less effective, we obtain
results comparable to their's. We believe that this excellent performance is
possible because GRA incorporates the essential ingredients which allow the
optimization to be effective on all scales. To give evidence of this, we now
show that GRA is also extremely effective on a very different problem.

{\it The spin glass problem (SGP)} --- Spin glasses have long been a subject
of intense study in statistical physics. One of the simplest spin glass
models is that of Edwards and Anderson~\cite{EdwardsAnderson75} in which
Ising spins ($S_i = \pm 1$) are placed on a lattice and the interactions are
between nearest neighbors only. The corresponding Hamiltonian is
\[H = - \sum_{\langle i,j \rangle} J_{ij} S_i S_j \]
where the $J_{ij}$ are quenched random variables with zero mean. For our
purpose here, the spin glass problem consists in finding the spin values
which minimize $H$. To find this minimum with a genetic algorithm approach,
we need the ``building blocks'' of good configurations. This time, simply
looking at the variables (spin orientations) which are shared between
parents is not effective since the energy is unchanged when all the spins
are flipped. Instead, we consider {\it correlations} among the spins. The
simplest correlation, whether two neighboring spins are parallel or
anti-parallel, will suit our needs just fine. Consider first any set of
spins; if the relative orientations of these spins are the same for all $k$
parents, we say that they form a ``pattern''; the values of the spins in
that pattern are then frozen up to an overall sign change. Now we sharpen a
bit this notion of a pattern: we require the set of spins to be both maximal
and connected, and we call such a set a block. (Note that the patterns
introduced for the TSP also had these two properties.) We can associate a
fictitious or ``blocked'' spin to each such block to describe its state.
Flipping this blocked spin corresponds to flipping all the spins in the
block, a transformation which maintains the pattern (i.e., the relative
orientations of the spins in the block).

With these definitions, it is not difficult to see that each spin belongs to
exactly one block (which may be of size $1$ though). Furthermore, the
configurations compatible with the patterns shared by the $k$ parents are
obtained by specifying orientations for each blocked spin; this procedure
defines the space spanned by all possible children. Not surprisingly, the
energy function (Hamiltonian) in this space is (up to an additive amount)
quadratic in the blocked spin values, so finding the {\it best} possible
child is again a spin glass problem, but with fewer spins! Because of this
property, the renormalization/recursion approach can be used very
effectively, similarly to what happened in the case of the TSP.

To find the (renormalized) coupling between two blocked spins, proceed as
follows. First put the two spins in the up state; unblock each spin so that
one has all the spins of the initial system they are composed of. The
coupling between the two blocked spins is obtained by summing the $J_{ij}
S_i S_j$ where $S_i$ belongs to the set defining the first spin and $S_j$ to
that of the second. (Here, $S_i$ denotes the value ($\pm 1$) of the spin $i$
when its (unique) blocked spin is up. Note also that to obtain the total
energy of a blocked configuration, one also has to take into account the
energy inside each blocked spin.) Finally, a straightforward calculation
shows that this formalism carries over in the presence of an arbitrary
magnetic field also.

Given the construction of blocks and a local search routine (we use a
version of the Kernighan-Lin~\cite{KernighanLin70} algorithm (KL)), the GRA
proceeds as before. For the number of parents $k$, we follow the spirit of
the procedure used for the TSP: we increase $k$ dynamically until the size
of the renormalized problem is at least 7.5~\% that of the current problem.
For this choice, $k=5$ is the most frequent value, and we find that the
distribution of $k$ is rather narrow. (Clearly, when $k$ increases, the size
of the renormalized problem increases rather rapidly.)

Testing the algorithm is not easy as there is no library of solved SGP
instances. Fortunately, when the grids are two-dimensional, there are very
effective exact methods for finding the optimum~\cite{SimoneDiehl96}. We
thus performed a first type of test where we ran our GRA on ten instances
corresponding to toroidal grids of size $50\times 50$ with $J_{ij}=\pm1$.
(The exact solutions were provided by J. Mitchell.) For these runs we set $M
= 5 + 0.2N$ for each level ($N$ being the number of spins at that level).
The algorithm halted on the 6th, 7th, or 8th generation, and in all cases
found the {\it exact optimum}. Furthermore, we measured the mean excess
above the optimum for each generation. The first generation corresponds to
simply using the local search, and had a mean excess above the optimum of
12~\%. Thereafter, the mean excess energy decreased by a factor of 2 to 3 at
each generation, until it hit 0. (Furthermore, instance to instance
fluctuations were small.) In terms of computation time, our local search
took on average 0.02 seconds on these instances, and the average time taken
by GRA was 16,000 seconds. This performance is competitive with that of the
state of the art heuristic algorithm~\cite{Hartmann99} developed
specifically for the SGP. Since this same property was found to be satisfied
in the case of the TSP, there is good evidence that GRA is a {\it general
purpose} and effective optimization strategy.

\begin{table}
\begin{center}
\begin{tabular}{cccccc}
$L$ & $\Delta_{KL}$ & $\tau_{KL}$ & $\Delta_{GRA}$
& $\tau_{GRA}/\tau_{KL}$ & $P_{GRA^*} $\\
\hline
4 & 8.2 \% & 7.9 $10^{-4}$ & 0.087 \% & 19 & 99.8 \%\\
6 & 11.5 \% & 1.5 $10^{-3}$ & 0.65 \% & 43 & 98.6 \%\\
8 & 13.5 \% & 2.7 $10^{-3}$ & 1.09 \% & 85 & 98.0 \%\\
10 & 14.1 \% & 7.1 $10^{-3}$ & 1.26 \% & 104 & 94.0 \%\\
\end{tabular}
\end{center}
\caption{Tests on $L\times L\times L$
SGP instances. $\Delta_{KL}$ and $\Delta_{GRA}$ are the relative
differences between the energy found by the corresponding algorithm and the
optimum, $\tau_{KL}$ and $\tau_{GRA}$ are the CPU times in seconds to treat
one instance. $\tau_{GRA}$ and $\Delta_{GRA}$ are results for $M=15$.
$P_{GRA^*}$ represents the probability for GRA to find the optimum when $M=N$
at the top level and $M=5+0.2N$ for inner levels.}
\label{TableSG}
\end{table}

As a second kind of check on our method, we considered $3$-dimensional grids
of size $L \times L \times L$ with Gaussian $J_{ij}$'s for which exact
methods are not so effective. These kinds of grids are of direct physical
relevance~\cite{Young98}. Since we did not know the exact optima, our
analysis relied on self-consistency: we considered we had found the optimum
when the most powerful version of our algorithm (large $M$) output the same
configuration with a probability above 90~\%. Measuring this probability
requires performing many runs, but once one has this putative optimum, one
can measure the performance of the algorithm in a quantitative way. To
achieve the precision required we set $M=N$ for the top level and $M=5+0.2N$
for inner levels; then the probabilities to find the optimum are as given in
the last column of Table~\ref{TableSG}. We also give in this table the
performance of the GRA with $M=15$ for {\it all} the levels; for this choice
of $M$, the algorithm is one to two orders of magnitude slower than KL, but
leads to mean energy excesses that are $10$ to $100$ times smaller! Overall,
the quality of the solutions is excellent even with a relatively small $M$,
and we see that up to 1000 spins, GRA is able to find the optimum with a
high probability provided $M$ is large enough.

\paragraph*{Discussion ---}
For both the traveling salesman and the spin glass problems, our genetic
renormalization algorithm finds solutions whose quality is far better than
those found by local search. In a more general context, our approach may be
considered as a systematic way to improve upon state of the art local
searches. A key to this good performance is the treatment of multiple scales
by renormalization and recursion. The use of a population of configurations
then allows us to self-consistently optimize the problem on all scales. Just
as in divide and conquer strategies, combinatorial complexity is handled by
considering a hierarchy of problems. But contrary to those strategies,
information in our system flows both from small scales to large scales and
back. Clearly such a flow is necessary as a choice or selection of a pattern
at small scales may be validated only at much larger scales.

In this work, we put such principles together in a simple manner;
nevertheless, the genetic renormalization algorithm we obtained compared
very well with the state of the art heuristics specially developed for the
problems investigated. Improvements in the population dynamics and in the
local search can make our approach even more powerful. We thus expect
genetic renormalization algorithms to become widely used in the near future,
both in fundamental~\cite{HoudayerMartin99b} and applied research.

\acknowledgments
We are grateful to O. Bohigas for his comments and to J. Mitchell for
providing the spin glass exact solutions. O. C. M. acknowledges support from
the Institut Universitaire de France. J. H. acknowledges support from the
M.E.N.E.S.R.

\bibliographystyle{prsty}
\bibliography{/tmp_mnt/home/houdayer/Papers/Biblio/references}

\end{document}